\documentclass[iop]{emulateapj}
\usepackage{epsfig}
\usepackage{amsmath}
\usepackage{natbib}

\newcommand{\lpp}{$T_{\rm LPP}$}
\newcommand{\loglpp}{$\log_{10}(T_{\rm LPP})$}
\newcommand{\kepler}{{\it Kepler}}

\begin{document}

\title{A Machine Learning Technique to Identify Transit Shaped Signals}
\author{Susan E. Thompson\altaffilmark{1,2}, Fergal Mullally\altaffilmark{2,3}, Jeff Coughlin\altaffilmark{2}, Jessie L. Christiansen\altaffilmark{4}, Christopher E. Henze\altaffilmark{3}, Michael R. Haas\altaffilmark{3} and Christopher J. Burke \altaffilmark{2,3}}

\altaffiltext{1}{{\it a.k.a.} Susan E. Mullally, NASA Ames Research Center, Moffett Field, CA 94035, USA, susan.e.thompson@nasa.gov}
\altaffiltext{2}{SETI Institute, 189 Bernardo Ave Suite 100, Mountain View, CA 94043, USA}
\altaffiltext{3}{NASA Ames Research Center, MS 244-30, Moffett Field, CA 94035, USA}
\altaffiltext{4}{NASA Exoplanet Science Institute, California Institute of Technology, M/S 100-22, 770 S. Wilson Ave, Pasadena, CA 91106}

\shorttitle{LPP Transit Metric}
\shortauthors{Thompson et al.}

\begin{abstract}
We describe a new metric that uses machine learning to determine if a periodic signal found in a photometric time series appears to be shaped like the signature of a transiting exoplanet.  This metric uses dimensionality reduction and k-nearest neighbors to determine whether a given signal is sufficiently similar to known transits in the same data set.  This metric is being used by the \kepler\ Robovetter to determine which signals should be part of the Q1-Q17 DR24 catalog of planetary candidates. The \kepler\ Mission reports roughly 20,000 potential transiting signals with each run of its pipeline, yet only a few thousand appear sufficiently transit shaped to be part of the catalog. The other signals tend to be variable stars and instrumental noise.  With this metric we are able to remove more than 90\% of the non-transiting signals while retaining more than 99\% of the known planet candidates. When tested with injected transits, less than 1\% are lost. This metric will enable the \kepler\ mission and future missions looking for transiting planets to rapidly and consistently find the best planetary candidates for follow-up and cataloging.

\end{abstract}

\section{Introduction}
As the size and complexity of astronomical data increases, the analysis of these data sets will need to become increasingly automated. When searching for transiting exoplanets, the first step is to test whether the light curve contains a train of transits. Many search algorithms exist to find these periodic events in a time-series of data \citep{TPS,Berta2012,Kovacs2002,Defay2001}, but the task of selecting which of those truly look transit-like is commonly performed by eye. Exoplanet surveys, such as {\it CoRoT}, \kepler, {\it TESS} and {\it PLATO} \citep{Auvergne2009,Koch2010,Ricker2014,plato}, have found, or will find, more signals in their data than can easily be examined consistently by a small team of people. While one solution is to enlist citizen scientists such as the planet-hunters \citep{Wang2013}, another solution is to develop more sophisticated metrics to cull-out those detections that do not look like transits.  

Here we discuss and implement a machine-learning technique to determine whether a signal looks like a transit.  This problem can be formulated as a dimensionality reduction and clustering problem.  The light curves contain thousands of points that are used to describe the shape of the feature.  However, only a few dimensions are needed to describe whether the signal has the steep ingress and egress of a transit, as well as a flat profile outside of the transit event. The trick is to characterize the light curves such that a dimensionality reduction routine clusters signals that look like transits in a region of parameter space separate from those that do not look like transits. 

Similar work was done by \citet{Matijevi2012} to characterize the eclipsing binaries in the \kepler\ time-series data. Eclipsing binary stars range from detached, with discrete transit-like dips, to over-contact binaries, that continually vary.  \citet{Matijevi2012} used an algorithm known as Local Linear Embedding \citep[LLE,][]{LLE2000} to reduce the number of dimensions of the folded light curves of the reported binary stars down to one dimension. The class of binary star was now mapped onto a continuum of values with detached binaries on one end and over-contact binaries on the other. 

The problem we solve in this paper is somewhat different because we attempt to separate transit-like events,  i.e. periodic v- or u- shaped variations in the light curve,  from all other periodic events found by the \kepler\ pipeline \citep{pipeline,TPS,DV}. The \kepler\ search for transiting planets \citep{TPS} returns transit-like signals as well as other periodic variations. The most common type of false alarm is sinusoidal variations likely caused by spots, pulsations, tidal binaries or contact binaries.  But the search also returns erratic signals likely due to instrumental effects, or events that contain no obvious signal at all.  

While these signals were removed by hand in previous \kepler\ planet candidate catalogs \citep{Borucki2011,Borucki2011b,Batalha2013, Burke2014,Rowe2015,Mullally2015}, the Data Release 24 (DR24) Kepler Objects of Interest (KOI) catalog will determine whether a signal is a planetary candidate  using an entirely algorithmic approach using a set of algorithms called the Robovetter \citep{Coughlin2015}. The reason for this is driven by the desire to measure accurate planetary occurrence rates \citep{Burke2015,Dressing2015,Batalha2014}.   In order to measure the sensitivity of the pipeline to finding planets, the entire search must be done repeatedly on both real and injected signals, demanding automation.

 The basic philosophy of the \kepler\ Robovetter is that it uses various metrics to decide if a signal in the data 1) is not transit-like (in shape or significance), 2) has a significant secondary event (an indication that it is an eclipsing binary), or 3) shows evidence of being due to a background eclipsing binary. To improve the ability of the Robovetter to evaluate the first item, we create a metric that tests whether the shape of the transit is similar to known transiting events.

We describe here a metric that uses the Locality Preserving Projections \citep[LPP,][]{LPP} dimensionality reduction and k-nearest neighbors to accomplish this task. While we describe the method as it specifically applies to the \kepler\  data, the same technique could easily be adapted to run on the result of any high duty-cycle transit search.   In \S\ref{s:data} we describe the \kepler\ data and the signals found by the pipeline.  In \S\ref{s:metric} we describe how we characterize our data, determine a training set, and calculate the LPP transit metric for the Q1--Q17 DR24 KOI catalog.  We further evaluate the performance by injecting transits in \S\ref{s:injection}. Finally, in \S\ref{s:discussion} we discuss the performance of the technique on the \kepler\ data. 



\section{The \kepler\ Data}
\label{s:data}
The \kepler\ spacecraft has collected 17 quarters of time series data. Each quarter is approximately 90 days and the cadence of the observations is approximately 29.4 minutes.  As a result \kepler\ has obtained as many as 70,000 brightness measurements of over 160,000 stars listed in the Kepler Input Catalog \citep[KIC,][]{kic} spanning 4 years \citep{DR24}. The Kepler pipeline reduces the data, creates light curves, and searches these light curves for periodic signals that may be consistent with a transit.  These signals are known as Threshold Crossing Events (TCEs) and are available at the NASA Exoplanet Science Institute (NExScI) archive\footnote{http://exoplanetarchive.ipac.caltech.edu/}\citep{nexsci2013}.  As part of DR24, 20,367 TCEs were discovered by the \kepler\ pipeline \citep{Seader2015}.

The Transit Planet Search (TPS) component of the Kepler pipeline performs the search for the transit signals. It does this by whitening the data and searching for significant detections at a large range of periods and using 14 different transit durations. See \citet{Seader2015} and references therein for more information on how TPS searches for transit signals.  Once a signal is found, it is sent to the Data Validation (DV) module of the Kepler pipeline where it is fit with a transit model and the in-transit points are removed.  The same search algorithm is run on the gapped light curve until no more signals are found.  In this way, up to 10 TCEs, can be found at different ephemerides on the same Kepler target.  All of the TCEs and the metrics calculated by these two pipeline modules are available at NExScI.   

The technique we describe here only relies on a few of the values calculated by TPS and DV.  We use the period, epoch and duration of the TCE as reported by DV, which is established by fitting the transit model of \citet{Mandel2002} to the signal. To find the TCE, TPS calculates a Multiple-Event Statistic (MES), which gives a measure of the significance of the detected TCE\footnote{MES is a measure of how correlated the data are to a sequence of evenly spaced transit pulses, normalized by the strength of the noise.} \citep{TPS}. TPS only searches down to a MES value of 7.1\footnote{A MES limit of 7.1  ensures only one false alarm due to white noise during the duration of the \kepler\ mission \citep{Jenkins2002}.}; typically these TCEs appear only marginally above the noise.  Sometimes DV fails to fit the signal, in these cases we revert to the original period and epoch found by TPS and use the pulse duration that was being used when the signal was originally found.  The MES is available regardless of whether the transit model converged.


From these TCEs the \kepler\ project creates a catalog of planet candidates and astrophysical false alarms (e.g. binary stars and background binary stars), known as the KOI catalog. For every catalog published to date  \citep{Mullally2015,Rowe2015,Burke2014,Batalha2013,Borucki2011}, all TCEs were examined by a team of astronomers to determine if the data associated with each TCE could potentially be due to a transiting planet (a process known as vetting).  Starting with the Q1-Q17 DR24 KOI catalog, this activity is being done by what is being called the ``Robovetter" \citep{Coughlin2015,CRV2015,Marshall2015}; all tests to determine whether a TCE is a planet candidate will be made by a computer algorithm.  This is not an entirely new concept; in previous catalogs some of the tests performed on the transit-like TCEs, specifically to determine whether a transit event occurs on the star in question, were performed by the ``Centroid Robovetter" \citep{CRV2015}, also \citet{McCauliff2015} implemented a machine learning approach to evaluate the \kepler\ TCEs. For the DR24 KOI catalog that same philosophy of using metrics and logic is being applied to evaluating the transit shape and whether the transit has a significant secondary eclipse (an indication that it is an eclipsing binary star).  The metric we have implemented here focuses on the first question, ``Does this TCE look transit-like?".


\section{Calculating the LPP Transit Metric}
\label{s:metric}
In this section we describe the procedure used to calculate the metric for the DR24 KOI catalog.  To summarize, we start with detrended light curves for each TCE. We then fold and bin each light curve into $N$ points.  These $N$ points act as the initial number of dimensions that describe each TCE.  Using a high signal-to-noise subset of these binned TCEs, we create a map from the initial $N$ dimensions down to a smaller $n$ dimensions using the LPP dimensionality reduction algorithm \citep{LPP}. That map is applied to all TCEs. Then, to find the area of this reduced dimensionality space where the transits lie, we create a labeled data set of known transit-like TCEs. For each TCE, we measure the average Euclidean distance to the $k$ nearest transit-like TCEs. This average is the value of the LPP transit metric for the TCE. Each of these steps are discussed in more detail below and the final values for the DR24 TCEs are given in Table~\ref{t:lpp}.

\newpage
\subsection{Detrending the Light Curves}
The \kepler\ vetting activity uses two different detrending algorithms to evaluate each TCE. Primarily it uses the harmonic-removed, median detrender calculated by the DV portion of the Kepler pipeline \citep{DV}.  In this case a harmonic series of the largest sine-waves are fit and removed, and then a median detrender is applied with the time scale selected by considering the duration of the signal. We refer to this detrender as the DV-median detrender.  The alternate detrender is a non-parametric penalized least-squared (LS) method from \citet{Garcia2010} which includes only the out-of-transit points when computing the filter. Because the in-transit points are not used when detrending, this detrender has the effect of making most signals look more like transits. We refer to this detrender as the penalized-LS detrender. In previous versions of the KOI catalog, members of the team would consider the folded and binned light curve produced by both the DV-median detrender and the penalized-LS detrender to determine if the signal looked sufficiently like a transit. More information about these detrenders and how they have been used to vet planet candidates can be found in \citet{Mullally2015} and \citet{Rowe2015}. After detrending the median value of the light curve is set to zero and the variations are given as fractional changes in the observed brightness.

\subsection{Feature Extraction}
In order for our metric to examine the shape of each possible transit found in the data, we start by folding the time series and centering the event at a phase of 0.5.  We then need to extract, or encode, the shape of this light curve into the same number of data points for all TCEs.  To use the dimensionality reduction technique we describe below, all TCEs must start with the same number of dimensions, i.e. data points.  Fitting would be an option, as done in \citet{Matijevi2012}. However, since we are trying to separate non-transit from transiting phenomena, it is not clear what sort of function would easily account for all the different types of variability found among the TCEs as well as a transit shape.

Instead we bin the folded light curve. The number of bins that optimally describes a light curve depends on the time scale of the interesting phenomena.  Most transits look like a rapid dip in the brightness of the star; the transit takes place for a relatively short amount of time compared to the full orbit. For instance, a typical duration for an object in a 20 day orbit is around 10~hours, covering only 2\% of the light curve. An Earth transiting a Sun takes $\approx$12 hours; a mere 0.14\% of the light curve would be in transit.  However, short-period, over-contact eclipsing binaries, stellar spots, and some short-period transiting planets can be fit with a duration lasting upward of 20\% of the folded light curve.  Since it is the shape of the detected dip in the light curve that we are interested in, and not the duration, we choose our bins in such a way that all transits span approximately the same number of binned points. This is done by using the period and transit duration when picking the range of phases to bin.  Also, for the relatively short duration transits, this method has the effect of increasing the importance of the in-transit phases when deciding if the event is transit-like.

However, those signals that confuse transit detectors will similarly fool this metric unless we also consider the light curve at phases away from the transit. For instance variable stars can look like a broad transit except when you also consider the variability at phases outside of the purported transit event. So, we also include bins at phases away from the transits, and include enough to be able to detect large variability at low harmonics of the detected period.

Finally, in order to encourage our algorithm to discriminate based on the shape of the light curve, and not the amplitude of the signal, we normalize all our depths to negative one, based on the lowest binned value during the transit.

To summarize, in order to prepare the input matrix to the LPP dimensionality algorithm we did the following. We started with detrended data, we fold on the TCE period, and then choose approximately one third of our binned points to cover the phases that lie 5 transit durations on either side of the reported event. The other two-thirds of our binned points are evenly spaced across phases that do not include those near 0, 0.5 and 1.  For the exact way the bins were chosen for the \kepler\ TCEs, see \S\ref{s:results}.  By not binning points near a phase of 0 we remove the effect of significant secondaries on a number of TCEs with little impact on the measurement of the out-of-transit light curve shape. Consequentially, those TCEs with significant secondaries will still look like transits. The \kepler\ Robovetter has other metrics in place to remove TCEs with significant secondaries.  We sort all of the binned values by phase and normalize all of the binned points such that the smallest measured binned value in the in-transit bins has a value of negative one. 

\begin{figure*}[h!]
\includegraphics[scale=0.68]{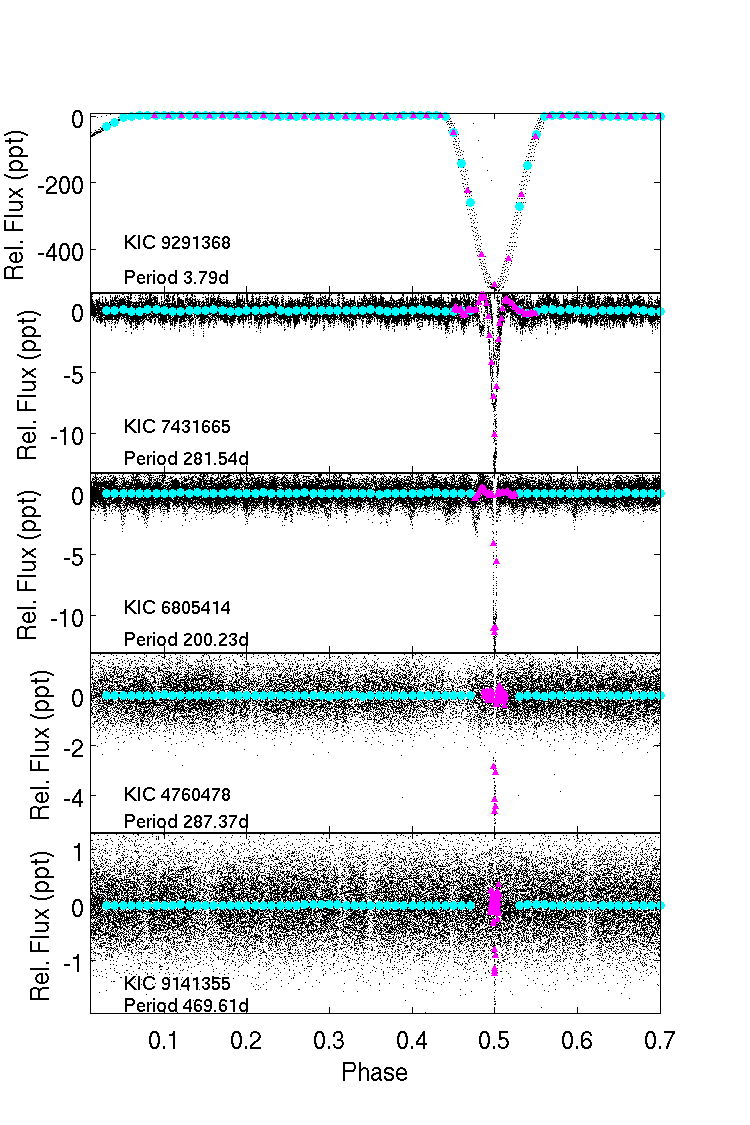}
\includegraphics[scale=0.68]{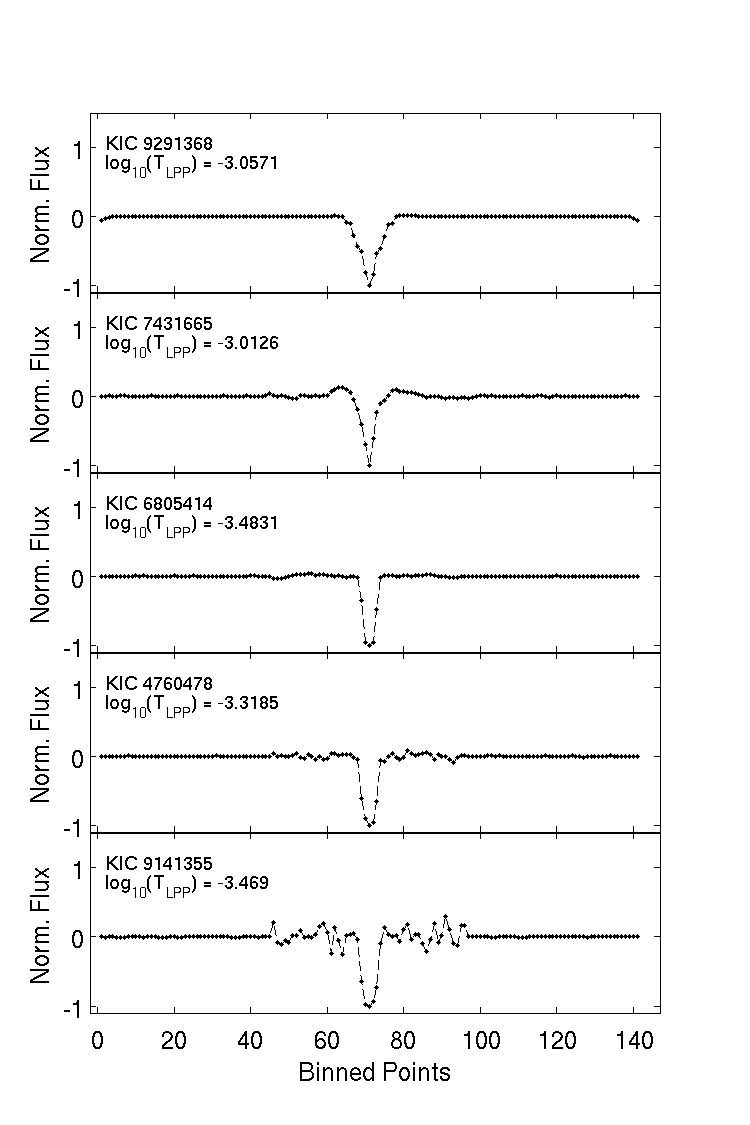}
\caption{\label{f:bintransit} Left: The folded light curves for five example TCEs (black) along with the binning in units of parts per thousand. Those binned points that span 5 times the transit duration are shown with magenta triangles and those that cover all but the in-transit points are shown with cyan circles. Right: The normalized binned points sent to the LPP algorithm.  This method of preparing the data has the effect of making all transit have the same depth and similar widths. Because the width of the in-transit bins can be much smaller, the binned points on the right can appear much noisier near the transit. \newline }
\end{figure*}

\begin{figure*}[h!]
\includegraphics[scale=0.68]{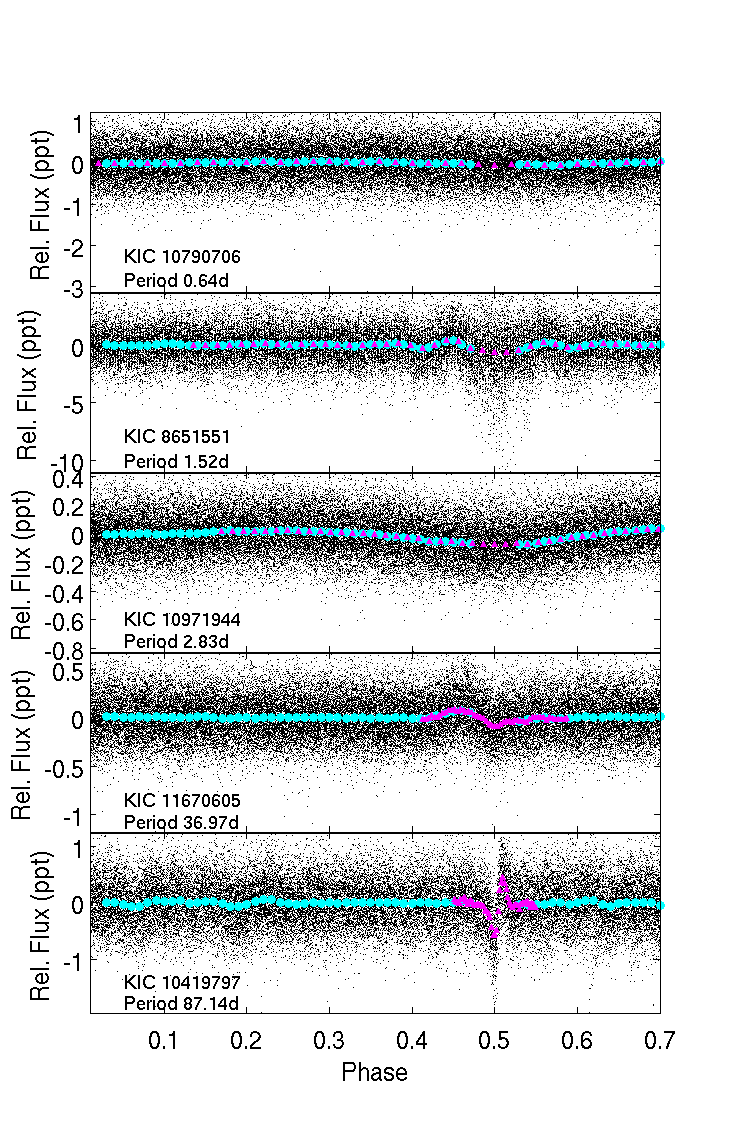}
\includegraphics[scale=0.68]{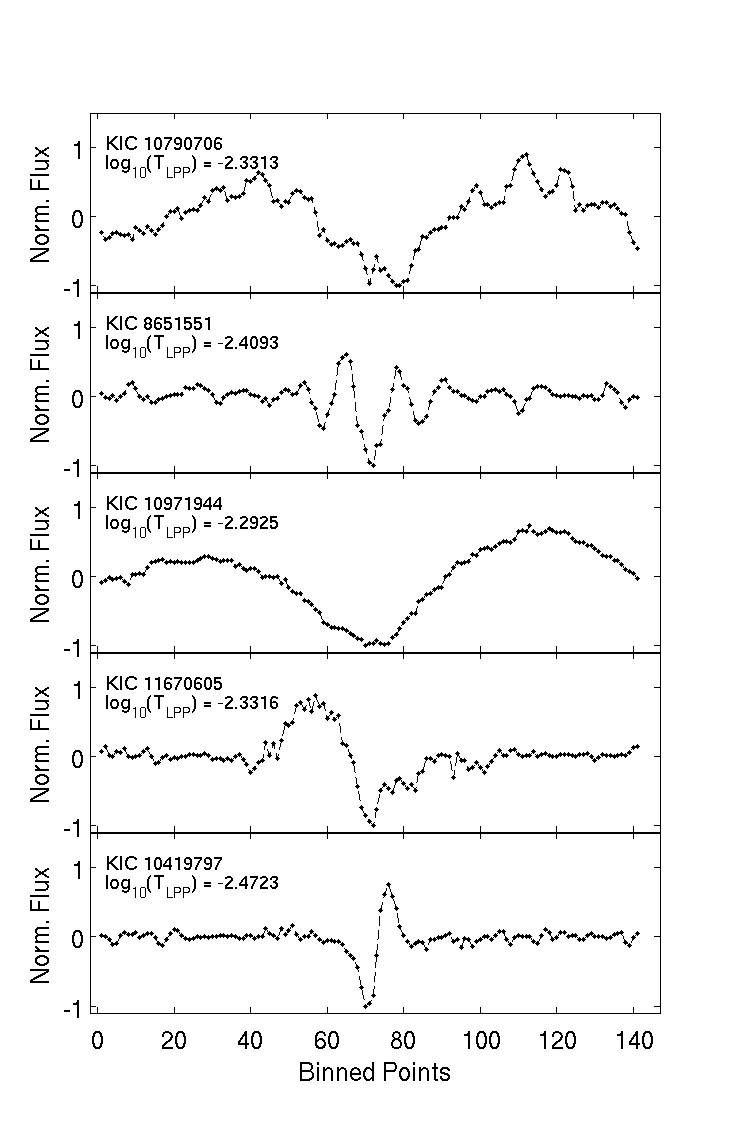}
\caption{\label{f:binvariable} Left: The folded light curves of TCEs that are likely not transiting planets (black) in units of parts per thousand. Both groups of binning are shown, those that span 5 times the transit duration are shown with magenta triangles, and those that cover all but the in-transit points are shown with cyan circles. Right: The normalized binned points sent to the LPP algorithm. \newline }
\end{figure*}

We show some examples of binned light curves with transits, using the DV-median detrender, in Figure~\ref{f:bintransit} and non-transiting binned light curves in Figure~\ref{f:binvariable}. Notice that for long-duration, short-period TCEs, the two sets of binned points overlap in phase.  For both figures the input to the dimensionality algorithm is shown on the right; the phases are sorted and the depth is normalized.  As a result of this binning procedure, the general shape of the transit is preserved but information about the depth and duration of the transit have been removed.  

\begin{figure*}[!h]
\includegraphics[scale=0.68]{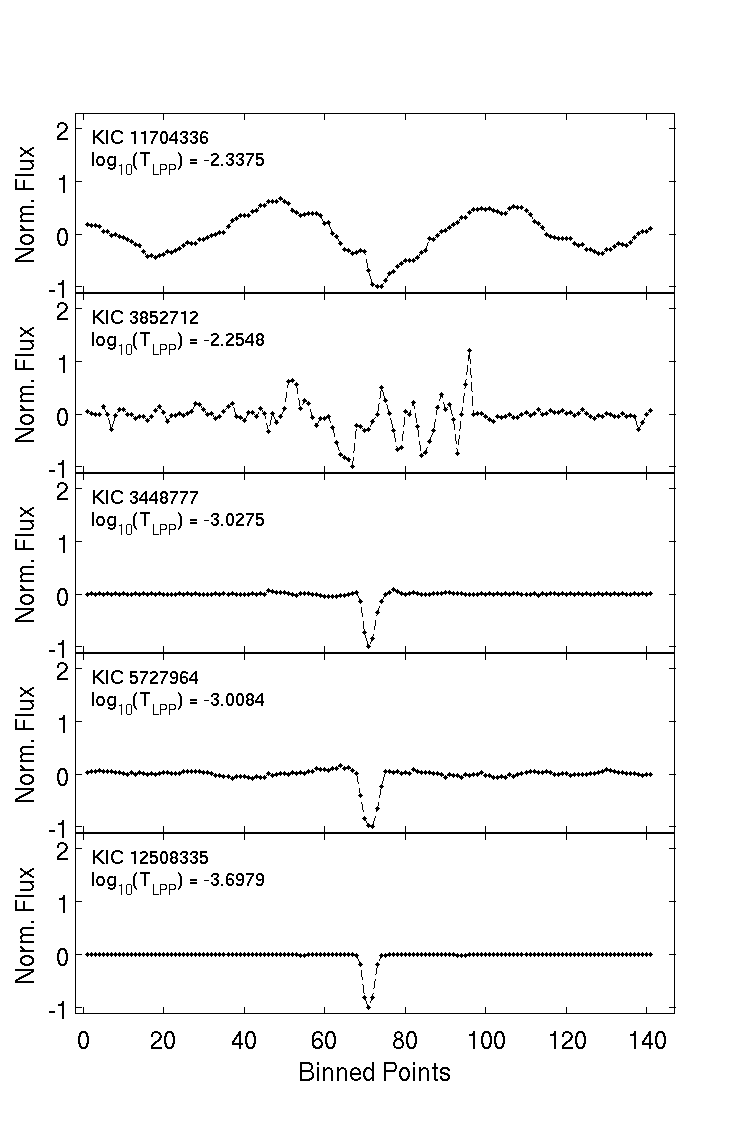}
\includegraphics[scale=0.68]{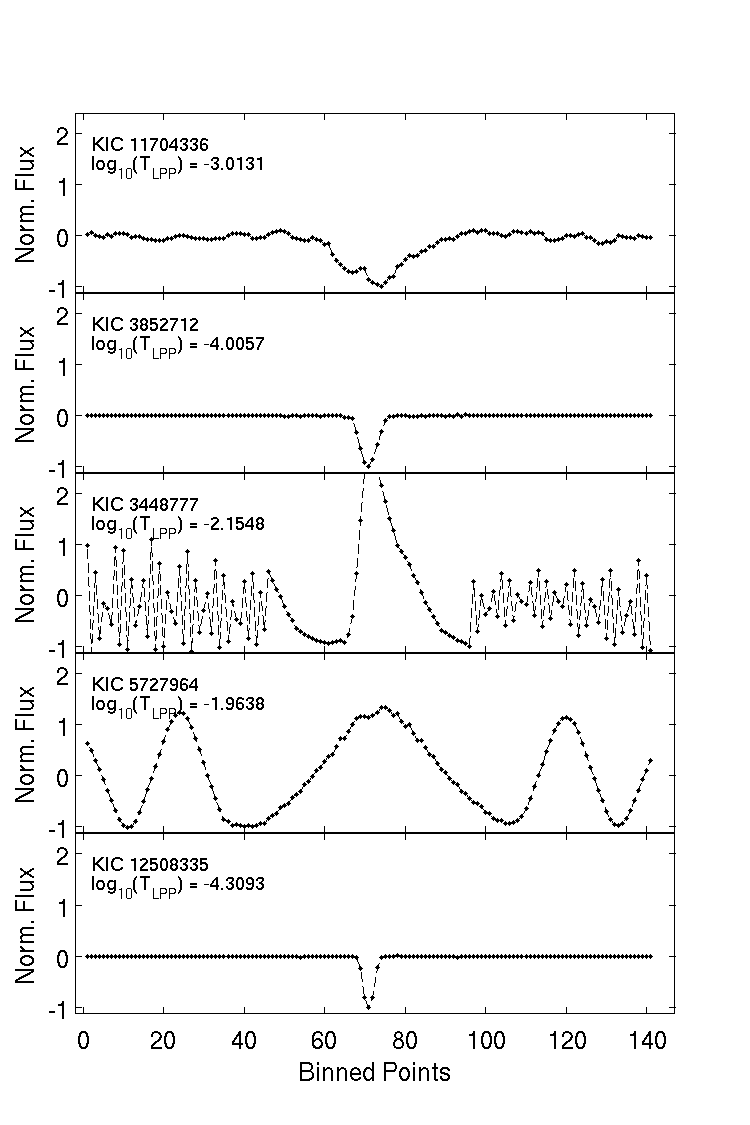}
\caption{\label{f:detrend} The binned points for the same TCEs as seen by the two different detrenders. The DV-median detrender is on the left and the penalized-LS detrender is on the right. The first four panels show TCEs where the detrenders disagree and the last case is an example where they agree. The \loglpp\ value is given on the plot below the KIC number. Typically, a value larger than -2.5 indicates a non-transit like shape. \newline }
\end{figure*}

To give an idea of how much the detrenders can disagree on the shape of a TCE, we show in Figure~\ref{f:detrend} the binned points for the same TCEs with the different detrenders. In the first four panels, we specifically pick cases where the detrenders disagree so much that it significantly changes the outcome of the LPP transit metric's value. In these cases the light curve is extremely variable; how much of that variability is removed by the detrender will drastically change the appearance of the light curve, especially when it is folded and binned. 

%

\subsection{Dimensionality Reduction}
Because we are trying to reduce the complex information of a binned light curve down to the simple question of whether it looks like a transit or not, we considered the power of dimensionality reduction. While many of these routines which attempt to maintain the most diverse dimensions in a data set could work for our purposes, we settled on Locality Preserving Projections (LPP)  /\footnote{http://papers.nips.cc/paper/2359-locality-preserving-projections.pdf}\citep{LPP} as implemented by the \textsc{matlab} Toolbox of Dimensionality Reduction\footnote{http://lvdmaaten.github.io/drtoolbox} \citep{matlabLPP}.  One advantage of LPP over nonlinear algorithms, such as Local Linear Embedding, is that the mapping is unambiguously defined everywhere in the higher dimensionality space; and so LPP can be precisely evaluated on injected signals, even though it was trained on only the original sample. 

LPP is similar to the most commonly used linear dimensionality reduction algorithm, Principle Component Analysis (PCA). While PCA attempts to maximize the variance of the data in each dimension, LPP attempts to preserve the local neighborhoods, as measured by a Euclidean k-nearest-neighbors, when reducing the dimensions.  In this way LPP has the advantages of being less sensitive to outliers than PCA and also has the ability to map highly nonlinear manifolds as can be done by nonlinear techniques. For more information on precisely how this algorithm works see  Section \ref{s:lpp}\ and  \citet{LPP}. While the linearity and locality aspects of LPP are useful in the problem we are trying to solve, ultimately we chose this particular dimensionality reduction algorithm because it did a good job of preserving the transit-shape information in binned light curves.  

The LPP algorithm takes our $N$ binned points for a sample of the TCEs, and maps them to a lower number of dimensions, $n$, attempting to preserve those elements of the set that are adjacent to each other, as defined by using k-nearest neighbors. Once our TCEs are represented by the lower number of dimensions we calculate the distance to an integer number, $k$, of nearest known transits from the labeled data set described below. We use \textsc{matlab}'s \emph{knnsearch} algorithm using a Euclidean distance (in \textsc{matlab} this is performed using a Minkowski distance with an exponent set to 2).  The mean of these $k$ distances is what we report as the LPP transit metric. For ease, we use the symbol \lpp\ to represent the LPP transit metric.  If the TCE is as close to the known transits as the known transits are to each other, the TCE is deemed to be transit-like.



\subsection{Creating a Labeled Data Set}
\label{s:label}
The key to the success of this metric is having a good set of known transit-like signals for training.  In this case we create a labeled data set from those TCEs previously found and vetted by the \kepler\ Project  \citep{Batalha2013,Burke2014,Rowe2015,Mullally2015}.  Those TCEs whose ephemeris matched those of known KOIs in the cumulative table at the NExScI archive (except for those marked with the 'not-transit-like' flag) became the sample of known transiting objects.  Note, KOIs that have dispositions of false positive due to being a centroid offset or because of an ephemeris match are also included as transit-like objects in our labeled data set.

For testing purposes, we also federated the TCE sample with those TCEs from earlier catalogs (specifically Q1--Q16 and Q1--Q12) that previously failed to become a KOI. These objects were all evaluated by individuals and deemed either to look not-transit-like or were too low signal-to-noise to be made into a KOI.  Those non-KOI objects that matched the ephemeris of our Q1-Q17 TCE list were labeled as not-transit like.  

We separately track two groups within the transit-like population. First, we track those objects that are known to be planetary candidates in the cumulative KOI table at NExScI. These make a very high fidelity population of objects known to have a transit shape.  Second, we track those objects with a known secondary eclipse, and likely to be eclipsing binaries by using the false positive flag available at NExScI. The intent is to keep the eclipsing binaries at this stage in the vetting because many of them look sufficiently like transiting planets. Other metrics are used by the \kepler\ Robovetter to remove these from the planet candidate sample. So, those objects with known significant secondaries are counted as transit-like.

In total we label 5678 TCEs as transit-like, of which 3738 are planetary candidates, 633 have a significant secondary, and 1307 are other known false positives, and we label 1039 as not-transit-like. The entire transit-like population is important because that is what defines the parameter space that should contain transit-like TCEs.  Note, we set aside 10\% of the transit-like set for testing purposes and did not use them to train our metric. Table~\ref{t:lpp} provides the labels given to each TCE for training and testing purposes.  

There are a few small issues with creating the labeled data set in this way. Some of the low signal-to-noise objects from previous TCE catalogs will now be legitimate transit-like signals because more transits are available and because of improvements to the pipeline that remove known noise sources. As a result, we might not expect all of the not-transit-like TCEs to fail our metric. Also, the false positive flags available at the archive were not universally set for every vetted KOI. The largest impact of this is that a very small population of not-transit-like KOIs are masquerading as transit-like objects (likely less than 1\% of the transit-like objects). Also, some known binaries may not be included in the significant secondary set. These are issues we keep in mind when using the labeled data set for training and testing our metric. 


\subsection{Evaluating the Q1--Q17 DR24 TCEs}
We apply the above technique to calculate \lpp\ for every Q1-Q17 DR24 TCE and give the values in Table~\ref{t:lpp}. While there are several tune-able parameters in creating the LPP transit metric (e.g. $k$, $N$, $n$, phase span of the in transit bins, etc.), we discovered that changing these parameters does not drastically change the outcome.  The values chosen for these parameters were determined empirically by trying to remove the most known non-transit like signals while keeping the most transit like signals. 

For this implementation, which is being used by the DR24 \kepler\ Robovetter \citep{Coughlin2015}, we started with $N=141$ binned data points, 51 in-transit and 90 out-of-transit. This was chosen so that the spacing of the out-of-transit bins is 0.01 in phase and the spacing of the in-transit bins is 0.2 times the duration in phase. The out-of-transit bins were evenly spaced across the phases 0.03--0.47 and 0.53--0.97. For most TCEs, this results in the bin spacing of the in-transit bins being smaller than the bin spacing of the out-of-transit bins.  We then used LPP to reduce the dimensions of the data set to $n=20$, where the TCE locality was determined by a k-nearest neighbor test with k$=15$. We used all TCEs with a MES$>8$ to create the mapping to the lower dimension, thereby removing those signals that were likely dominated by noise. For every TCE we reduce the dimensionality with this mapping and measure the distance to the $k=15$ closest known transits. We provide the \lpp\ values for both detrenders in Table~\ref{t:lpp}.

\begin{figure*}[!h]
\includegraphics[scale=0.44]{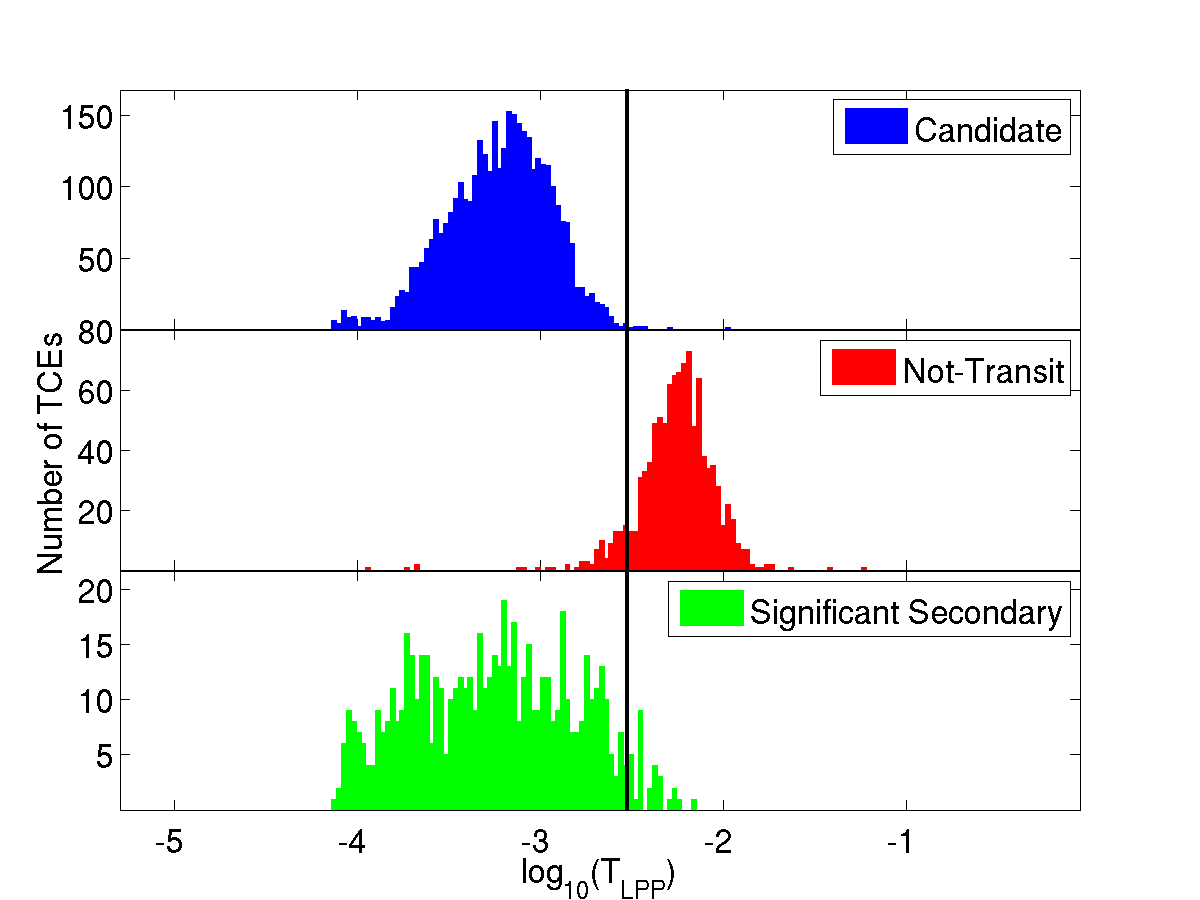}
\includegraphics[scale=0.44]{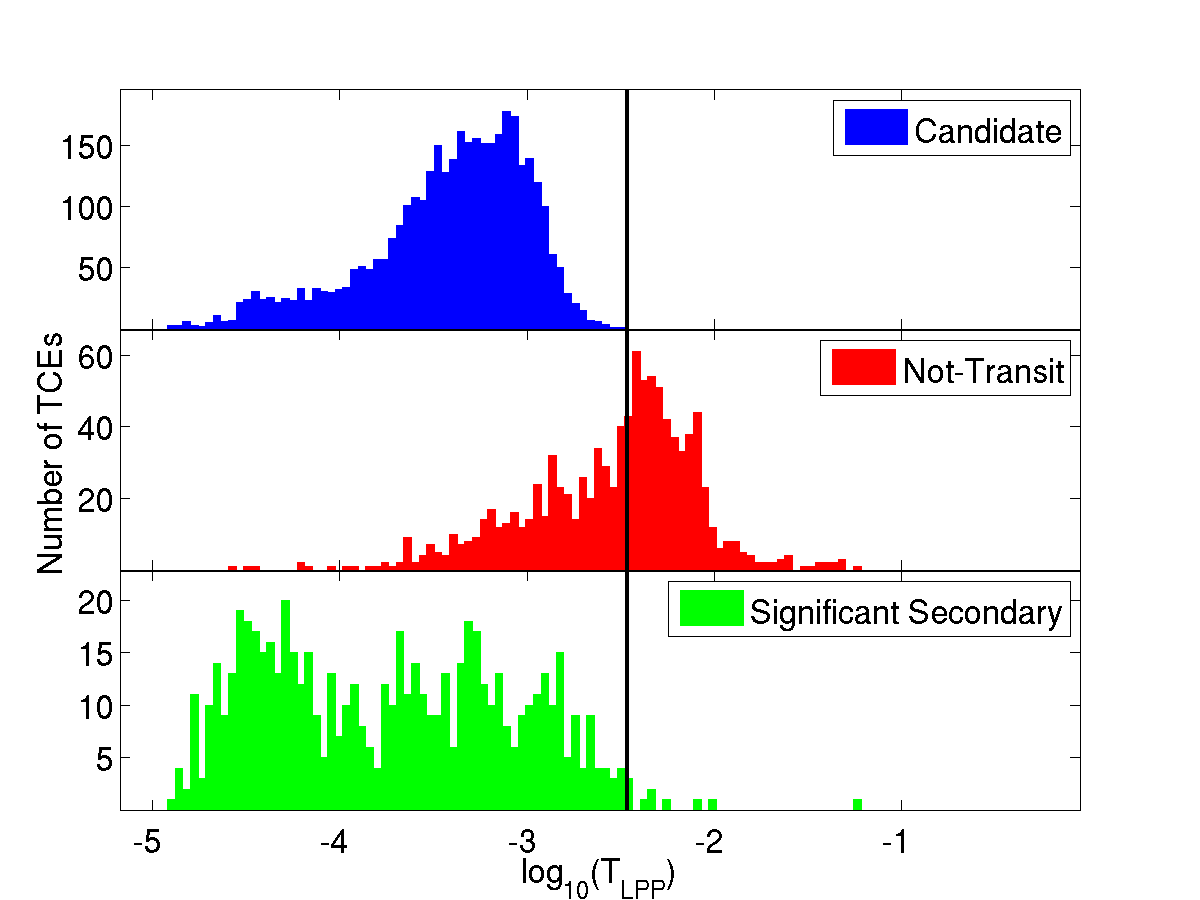}
\caption{\label{f:hist} A Histogram of the LPP transit metric for the labeled data set using the DV median (left) and penalized-LS (right) detrenders. A three sigma line discussed in the text is shown in black to help guide the eye.  Note, the transit-like population has been split into candidates and significant secondaries to show the slightly higher failure rate among binaries. \newline }
\end{figure*}

\label{s:results}While 20 dimensions are hard to display simultaneously, we can show the final metric and how it performed on the labeled TCEs. Figure~\ref{f:hist} shows the histogram of \loglpp\ calculated for the labeled data set.  There is significant separation in those objects that are known planet candidates from those that appear not-transit-like for the DV-median detrended data. One way to naively pick a value that could act as a cutoff line between transit-like and not-transit-like would be three times the standard deviation of \lpp\ for the known transit-like distribution. For the DV-median detrender this line would be \lpp$=0.003001$ (\loglpp$=-2.52$). If this line is chosen, \lpp\ rejects 0.3\% of the candidates, 4.4\% of those with significant secondaries, and 91.2\% of the not-transit-like objects. For the test set, the metric fails $<$0.1\% and 6.7\% of the candidate and binary-like sets, respectively. For the unlabeled data set, using this cutoff rejects 75.1\% of the TCEs.  

We show the same histograms for the light curves detrended by the penalized-LS in Figure~\ref{f:hist}. Notice the much poorer separation between the candidates and not-transit-like data sets. This detrender preserves the location of the known transit when detrending and as a result many variable star signals look like transits.  If we draw the same three sigma line based on the distribution of known transit-like events (\lpp$=0.00344$, \loglpp$=-2.463$), we reject only 51.7\% of the known not-transit-like TCEs while preserving $>$99.9\% of the transit-like and 98.6\% of the binary TCEs. We get almost identical values for the test set.  As a result, the penalized-LS \lpp\ value should only be used to supplement the results of the DV-median \lpp\ value, especially if the goal is to remove the most false signals from the catalog.   

\begin{deluxetable*}{crrrrrl}
\tablecolumns{7}
\tablewidth{0pc}
\tablecaption{\label{t:lpp}LPP Transit Metric for DR24 TCEs}
\tablehead{
\colhead{TCE}  &  \colhead{Period} & \colhead{Duration} & \colhead{MES}
& \colhead{\lpp} & \colhead{\lpp} & Label\\
\colhead{(KIC-num)} & \colhead{(days)} & \colhead{(hrs)} & \colhead{} &\colhead{DV-median} &\colhead{penalized-LS} & \colhead{}}\\

\startdata

000757450-01 &  8.88492 & 2.08 & 524.0 & 0.000237 & 0.000041 & TL-C\\
000892667-01 &  2.26211 & 7.51 &   8.0 & 0.004608 & 0.001884 & UNK\\
000892772-01 &  5.09260 & 3.40 &  15.6 & 0.001337 & 0.001081 & TL\\
001026032-01 &  8.46044 & 4.80 & 3888.7 & 0.000466 & 0.000083 & UNK\\
001026032-02 &  4.23022 & 4.61 & 1439.8 & 0.000303 & 0.000039 & UNK\\
001292087-02 &  1.09524 & 2.12 &   9.3 & 0.002109 & 0.004084 & NOT\\
001432214-01 & 161.78830 & 5.30 & 839.4 & 0.000235 & 0.000038 & TL-SS\\
001432214-02 & 161.77788 & 7.60 &  11.4 & 0.001103 & 0.001194 & UNK\\
001434660-02 &  0.52850 & 1.04 &  15.3 & 0.009662 & 0.005679 & NOT\\
\enddata
\tablecomments{The TCE period and duration are given for reference, see the NExScI TCE table for full precision of these values. The labels given to the data for training and testing can be interpreted with the following key, see \S\ref{s:label}: TL --  transit-like, C -- candidate, SS -- significant secondary, NOT -- not-transit-like, UNK -- unknown. Table 1 is published in its entirety in the electronic edition of the Astrophysical Journal. A portion is shown here for guidance regarding its form and content.
}
\end{deluxetable*}

\subsubsection{Code Availability}
The \textsc{matlab} code, as it was run to produce the LPP transit metric for the Q1--Q17 DR24 TCE list and the transit injection run, is available on SourceForge at http://sourceforge.net/projects/lpptransitlikemetric. It includes the code to bin the light curves and create the transit metric from the training set.

\subsection{Transit Injection}
\label{s:injection}
As a second test of the LPP transit metric, we apply it to transits that are injected into the light curves. This pixel-level transit injection is performed in order to understand what transit signals are not found by the \kepler\ pipeline, a necessary step to calculate the occurrence rate of planets \citep{Christiansen2013}. The transits were injected into all 17 quarters of data, but not with the expected distribution of the detected planets. Instead, they were injected with longer periods and at lower MES to carefully probe the parameter space where the \kepler\ pipeline is less likely to find planets. See \citet{Christiansen2013} for more information on transit injection for the \kepler\ pipeline. We used a 17 quarter transit injection run to test our metric. It produced 11,326 injected TCEs and used the same procedures as that discussed in \citet{TrInjection} and \citet{Christiansen2013}, but this is not the final transit injection run being used to calculate the average detection probability or test the Robovetter \citep{Christiansen2015,Coughlin2015}. This preliminary run was sufficient for our purposes of showing how well the metric preserves known transiting phenomena.


 
\begin{figure*}[]
\includegraphics[scale=0.44]{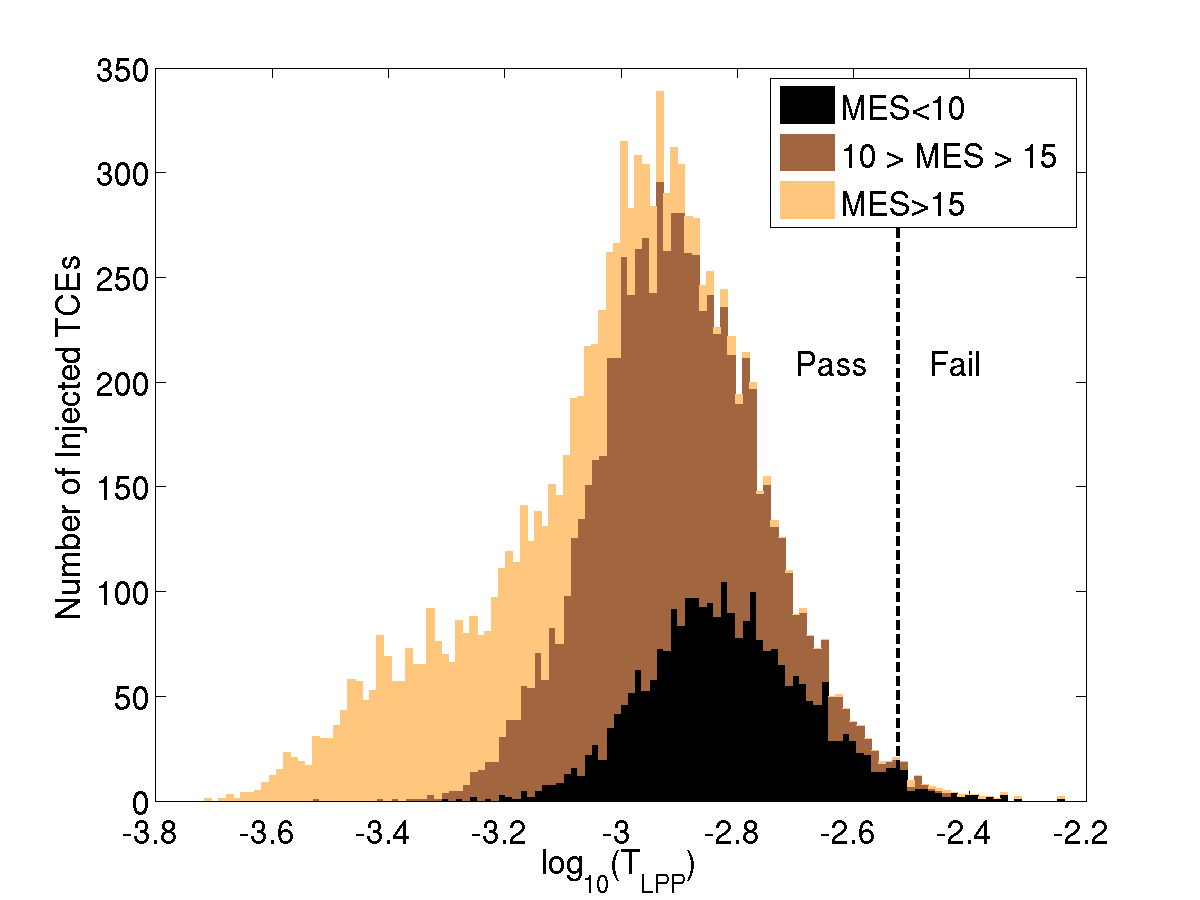}
\includegraphics[scale=0.44]{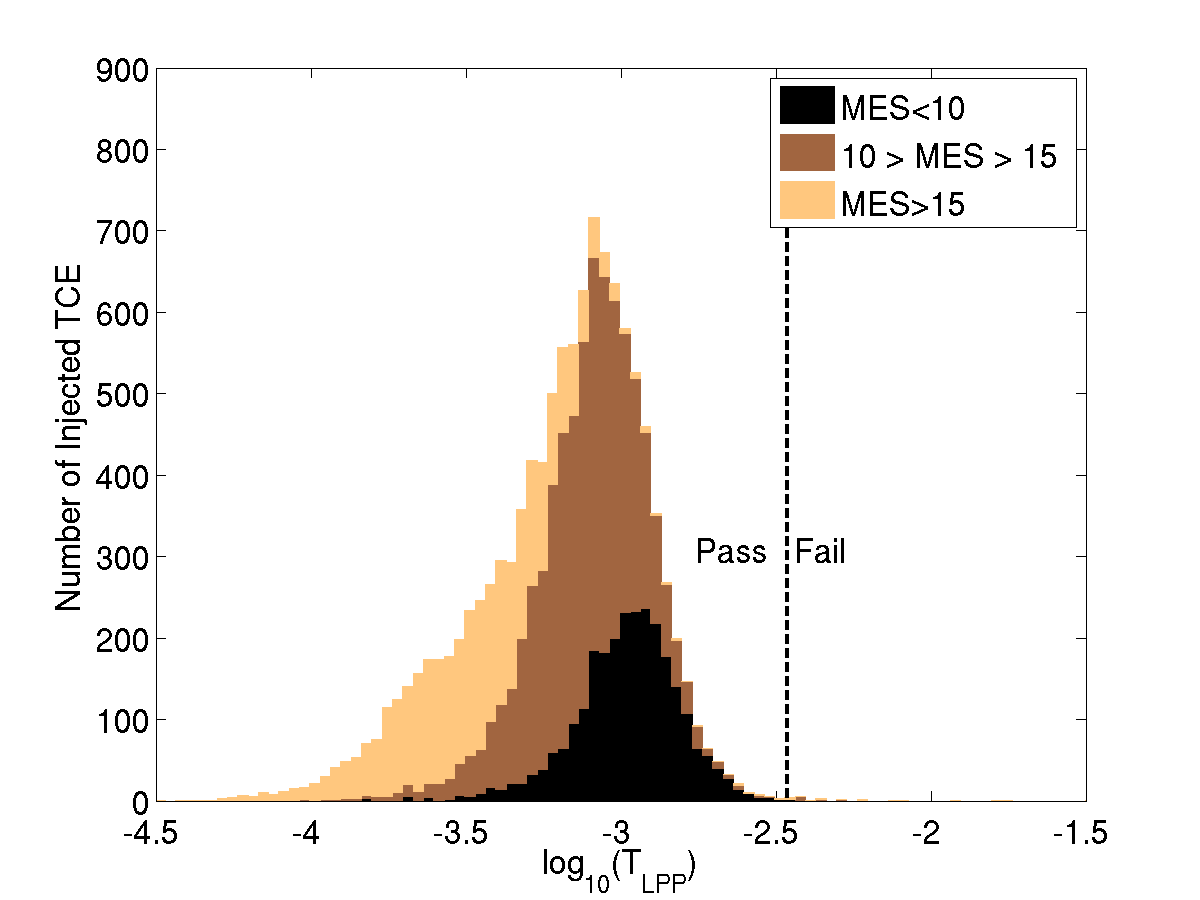}
\caption{\label{f:injectedHist} A Histogram of the LPP transit metric for the injected transits using the DV-median (left) and penalized-LS (right) detrenders. The histogram was calculated and stacked for different ranges of measured MES. Using the cutoff established by the real TCEs of log(\lpp)=-2.5, only a fraction of a percent of injected TCEs would be lost. \newline }
\end{figure*}

We use the procedure above to calculate \lpp\ for the injected transits using both detrenders. As stated before, we can easily apply our previous map (created from the original real TCEs) to the injected TCEs. Similarly when we apply the k-nearest neighbors, we are measuring the distance to the original set of known transit-like TCEs, not to the injected TCEs.  In Figure~\ref{f:injectedHist}, we show a histogram of the LPP transit metric for the 11,326 injected transits for both sets of detrended light curves. Notice that in general the lower MES transits have larger values for \lpp, meaning they are less well-separated from the non-transiting phenomena. This indicates that users of this metric may want to use a pass/fail threshold value that is dependent on MES in order to eliminate a larger fraction of the higher MES not-transit-like events without removing more border-line, low-MES events.  

Using the same value established above to indicate the line between transit-like and not-transit-like, we reject 94, or 0.8\%, of the injected transits for the DV-detrender and 23, or 0.2\%, of the injected transits for the penalized-LS detrending.  This is a slightly higher, though entirely acceptable, rate of failure for this metric when compared to the training set statistics given in Section \ref{s:results}.  

When inspecting those injected transits incorrectly classified by the DV-median metric as not-transit-like, we see that they are predominantly of low MES; 2\% of injected TCEs with recovered MES$<$10 fail the metric. Since low MES objects are predominantly transits of small radii, this introduces a very slight bias toward incorrectly classifying the signals of some of the smallest planets. This is not true for the penalized-LS detrender, where the minimal tail is distributed more evenly across MES.  Some of the higher MES injections that fail the metric are caused by injections onto highly variable stars and the detrenders did not preserve the shape of the transit. 

Also, there is no trend between the fraction of rejected signals as a function of period, even as the threshold is relaxed.  A larger investigation of the detection efficiency across various parameters using the entire Robovetter is available in the DR24 KOI catalog paper \citep{Coughlin2015} and with the detection efficiency products available at NExScI \citep{TrInjectionTcert}.


\section{Discussion}
\label{s:discussion}
The LPP transits metric is being used by the DR24 \kepler\ Robovetter \citep{Coughlin2015} to decide whether a TCE is a planetary candidate. The Robovetter is responsible for choosing a pass/fail threshold. In the past these catalogs have erred on the side of keeping questionable transits at the expense of retaining more false alarms. Even when this metric is implemented with a single, conservative value, as demonstrated above, it eliminates a large fraction of not-transit-like TCEs. Primarily, it does a good job of removing the short-period, long-duration TCEs that look like sine-waves, not transits.  The final results from the \kepler\ Robovetter and how it uses all of its metrics, including this one, will be available at NExScI.


Another automated transit finding technique is known as the Autovetter \citep{McCauliff2015,Jenkins2014}.  The Autovetter uses a random forest, machine-learning technique to decide which TCEs are planetary candidates.  For the Q1--Q17 DR24 TCE autovetter catalog \citep{Catanzarite2015}, the inputs now include the DV-median LPP transit metric. In this case, the random forest uses a training set (which is created in a manner very similar to how we created our training set) to learn how important \lpp\ is when deciding on whether a TCE is a transiting event.    The full results of the Autovetter's vetting will be found in the TCE table at NExScI for Q1--Q17 DR24. For this run, the LPP transit metric is ranked as the second most important metric for making its decisions \citep{Catanzarite2015}.

While there are several ways that we can change the value of \lpp\ for each TCE, by far the most important is how the data is prepared before applying the LPP dimensionality reduction. The number of chosen nearest neighbors and the number of reduced dimensions are relatively inconsequential in comparison to how the data is binned and how the data is detrended. This is exemplified in the performance for two different detrenders; the DV-median detrender does a far better job of preserving the differences between the transit-like and not-transit-like light curves. However, there are times when the penalized-LS detrending is more accurate. 

For example, one known issue with using this metric occurs for transits found on variable stars, especially if that variability is near to a harmonic of the period of the transit. These can be classified as not-transit-like because the detrending significantly distorts the signal.  Note, this effect could similarly fool the astronomers who did the manual vetting in previous catalogs. This is the reason two different detrendings were made available for the vetting activity. In those cases where the harmonic removal step of the DV-median detrender removed or distorted the transit, usually the penalized-LS would preserve it.  One way to improve our results would be to add some intelligence to the \kepler\ Robovetter to help it pick the more accurate detrender.

There are several populations of binary systems that are known to have higher failure rates with this metric. Eclipsing binaries with large eccentricities and a large secondary eclipse get flagged as not-transit-like because of the large deviation from a flat continuum at out-of-transit phases. Also eclipsing binaries make-up a smaller portion of the transit-like training set, and thus, it is more difficult for the deeper binaries to have 15 nearby neighbors. This problem could be mitigated by using the \kepler\ Eclisping Binary Catalog to add binaries to the transit-like training set \citep{Slawson2011,Prsa2011}. Finally, the Heartbeat stars, a class of dynamically distorted binary systems \citep{Thompson2012}, are classified as transit-like at times because the signal can include a large, discrete negative deviation in the brightness.  However, Heartbeat stars are rare ($<160$ are known in the \kepler\ field) and most known Heartbeat stars do not create TCEs ($<$10\% in Q1--Q17 DR24).

\section{Conclusions}
We present a new metric to discriminate between signals found by the \kepler\ Pipeline that look like transits and those that do not.  By folding and binning the light curve and applying machine learning techniques that rely on our knowledge about what a transit signal looks like in the \kepler\ data, we calculate the LPP transit metric. This metric is able to remove over 90\% of the known not-transit-like TCEs while preserving over 99\% of the known planet candidates and over 99\% of injected transit signals.  As currently implemented, this metric will prevent hundreds of not-transit-like events from populating the Q1--Q17 DR24 KOI table at the expense of losing only a few real transiting events. The Robovetter has additional metrics that will weed out some of the remaining signals, improving the reliability of the final catalog.

This LPP transit metric makes its decisions on TCEs consistently, reliably and rapidly, a large improvement over the manual activity performed by teams of astronomers in the past.  Because \kepler\ plans to calculate planetary occurrence rates, it is necessary to have an estimate of the planet catalog's completeness and reliability.  Using metrics like these, the Robovetter can quickly create a new catalog after any improvement to the pipeline, and also can run on injected signals. The latter makes it possible to more easily calculate the sensitivity of the pipeline across planet type for occurrence rate calculations.  

 Such a metric can easily be extended to work with current missions such as {\it K2} \citep{Howell2014,FM2015}, and future exoplanet missions such as the Transiting Exoplanet Survey Satellite mission \citep[TESS,][]{Ricker2014} and the Planetary Transits and Oscillations of Stars mission \citep[PLATO,][]{plato} to quickly identify the best transiting planets.  Metrics that can quickly and consistently evaluate potential signals in the data are even more important for these future missions because the data volume is even larger than \kepler\ and rapid vetting of the detected signals will enable immediate follow-up observations of the best planet candidates.


\begin{acknowledgments}
We thank the larger \kepler\ team for their support and hard work in making this data available and in supporting this paper. Funding for the \kepler\ mission is provided by the NASA Science Mission Directorate. We also thank the referee for the useful and insightful comments that improved the clarity of the manuscript. Some of the data presented in this paper were obtained from the Multi-mission Archive at the Space Telescope Science Institute (MAST). STScI is operated by the Association of Universities for Research in Astronomy, Inc., under NASA contract NAS5-26555. Support for MAST for non-HST data is provided by the NASA Office of Space Science via grant NNX09AF08G and by other grants and contracts.  This research has made use of the NASA Exoplanet Archive, which is operated by the California Institute of Technology, under contract with the National Aeronautics and Space Administration under the Exoplanet Exploration Program.\\

\end{acknowledgments}

\bibliography{KeplerPlanets}
\appendix
\newpage

\section{Locality Preserving Projections: A Summary of the Algorithm}
\label{s:lpp}
Here we summarize the basic logic of how LPP calculates its transformation matrix. For a more mathematically complete description of the LPP algorithm consult \citet{LPP}.  
 
The idea of LPP is to map a data vector $x_i$ (i.e. a folded, binned light curve) with $N$ dimensions to $y_i$ with $n$ reduced dimensions, i.e. $y_i = A^{\sc T} x_i$, where $A^{\sc T}$ is an $n$ x $N$ transformation matrix. $A$ is a matrix of eigenvectors calculated by minimizing the distance between reduced dimension vectors, $y$, only if they were near each other in the higher dimensions. 

LPP considers only the nearest neighbors in its calculations by constructing a symmetric weighting matrix, $W_{ij}$, that has a value of one when two data points are connected and zero otherwise. Data vectors $i$ and $j$ are considered connected to each other if $i$ is among the $k$ nearest neighbors of $j$ or if $j$ is among the $k$ nearest neighbors of $i$. In this way the number of connections for any one vector must be at least $k$, but can be larger when other vectors also consider $i$ a nearest neighbor. Using this weighting matrix, the function that is {\bf minimized} when choosing the eigenvectors, is as follows:
\begin{equation}
\sum\limits_{ij} (y_i - y_j)^2 W_{ij}
\end{equation}
Here, $i$ and $j$ run over the full number of data vectors (i.e. binned TCE light curves) in the data set. Because $W_{ij}$ only contains 1's when two $x$ vectors are adjacent, this function will be large when similar looking binned light curves are mapped far from each other. But non-neighboring vectors do not factor into the minimization. Thus, when LPP minimizes the above equation to find the transformation matrix, it preserves the local structure of the data. The eigenvectors with the smallest eigenvalues ($A$) are used as the transformation matrix to map $x_{i}$ to $y_{i}$. This transformation matrix may then be applied to any vector in the higher dimensions to reduce its dimensions. 

In comparison, PCA {\bf maximizes} the variance in the mapped data using the function:
\begin{equation}
\sum\limits_{i} (y_i - \bar{y})^2
\end{equation}
Because LPP only considers local distances, outliers will not weigh as heavily into the calculation and thus LPP can be more robust to outliers than PCA.  Also, because it tries to keep similar data vectors together, it can have better discriminating power than PCA, especially when the data vectors cluster into distinct groups in the higher dimensionality space.

\end{document}